\documentclass[preprint]{ptephy_om}
\preprintnumber{RIKEN-iTHEMS-Report-25}

\usepackage{amsthm}
\usepackage{bm}
\usepackage{booktabs}

\numberwithin{equation}{section}

\newtheorem*{thm*}{Theorem}
\newtheorem*{prop*}{Proposition}

\usepackage{graphicx,xcolor,tikz}
\usetikzlibrary{decorations.pathmorphing}
\tikzset{snake it/.style={decorate, decoration=snake}}

\newcommand{\magenta}[1]{#1}

\DeclareMathOperator{\res}{\mathrm{Res}}

\usepackage{hyperref}
\usepackage{orcidlink}

\title{Exact WKB method for radial Schr\"odinger equation}

\author[1]{Okuto Morikawa \orcidlink{0000-0002-0044-4491}\thanks{Corresponding Author}}
\affil[1]{Center for Interdisciplinary Theoretical and Mathematical Sciences (iTHEMS),
RIKEN, Wako 351-0198, Japan
\email{okuto.morikawa@riken.jp}}

\author[2]{Shoya Ogawa \orcidlink{0000-0003-0900-2486}}
\affil[2]{Department of Physics, Kyushu University, 744 Motooka, Nishi-ku,
Fukuoka 819-0395, Japan
\email{ogawa.shoya.615@m.kyushu-u.ac.jp}}

\begin{document}
\begin{abstract}
We revisit exact WKB quantization for radial Schr\"odinger problems from the modern resurgence perspective, with emphasis on how ``physically meaningful'' quantization paths should be chosen and interpreted.
Using connection formulae at simple turning points and at regular singular points, we show that the nontrivial-cycle data give the spectrum.
In particular, for the $3$-dimensional harmonic oscillator and the $3$-dimensional Coulomb potential, we explicitly compute a closed contour which starts at $+\infty$, bulges into the $r<0$ sector to encircle the origin, and returns to $+\infty$. Also we propose that the appropriate slice of the closed path provides a physical local basis at $r=0$, which is used by an origin-to-$\infty$ open path.
Via the change of variables $r=e^x$ ($x\in(-\infty,\infty)$), the origin data are pushed to the boundary condition of convergence at $x\to-\infty$, which renders the equivalence between open-connection and closed-cycle quantization transparent.
The Maslov contribution from the regular singularity is incorporated either as a small-circle monodromy which is justified in terms of renormalization group, or, equivalently, as a boundary phase; we also develop an optimized/variational perturbation theory on exact WKB.
Our analysis clarifies, in radial settings, how mathematical monodromy data and physical boundary conditions dovetail, thereby addressing recent debates on path choices in resurgence-based quantization.
\end{abstract}
\maketitle

\section{Introduction}
In recent years, the confluence of Borel analysis and resurgence theory has revitalized the exact WKB approach across a broad spectrum of physics, ranging from quantum-mechanical bound states to black–hole perturbation theory and quasi-normal modes (QNMs).
Resurgence provides a unified framework for handling Stokes analysis, monodromy, and Voros data behind formal asymptotic expansions~\cite{Voros:1983xx,Delabaere:1999xx,Iwaki:2014vad}, thereby rendering quantization as a consistency condition between Borel-summed local solutions and connection matrices~\cite{Sueishi:2020rug,Sueishi:2021xti,Kamata:2021jrs,Kamata:2023opn,Kamata:2024tyb}.
Early scattering–theoretic analyses of Schwarzschild QNMs, in particular Andersson~\cite{Andersson:1995zk}, already emphasized boundary-to-boundary formulations along the real axis.
Notably, contemporary advances in QNM analysis~\magenta{\cite{Miyachi:2025ptm,Hatsuda:2019eoj,Hatsuda:2021gtn,Matyjasek:2019eeu,Decanini:2011eh,Alfaro:2024tdr,Motl:2003cd,Hatsuda:2023geo,Kubota:2025hjk,DeAmicis:2025xuh,Yang:2024vor} (see also Refs.~\cite{Berti:2025hly,Mitman:2025hgy,Motohashi:2024fwt,Oshita:2025ibu,Oshita:2024wgt,Rosato:2024arw,Baibhav:2023clw,Cheung:2023vki})} demonstrate that one can formulate the spectral condition as an \textit{open connection problem} along a physically meaningful real-axis path (ingoing at the horizon and outgoing at spatial infinity), provided one incorporates logarithmic spiral Stokes curves and lateral Borel summation with due care. \magenta{In the context of quantum-mechanical resonances, we usually impose the boundary condition such that there exist no incoming waves at the asymptotic region, called the Siegert boundary condition~\cite{Siegert:1939}.}
This success, however, has also sharpened a conceptual tension: which path---or, more precisely, which homology class---should be deemed ``physical,'' especially when bound-state problems are traditionally quantized by closed cycles that count action areas, whereas scattering/QNM problems are posed by boundary-to-boundary zero conditions?

This paper addresses that tension for the radial equations derived from the $3$-dimensional ($3$D) Schr\"odinger equation.
Our viewpoint is that, while mathematics (in particular \magenta{the Aoki--Kawai--Takei (AKT) school of exact WKB~\cite{AokiKawaiTakei1991RIMS853,Kawai:2005} and seminal works~\cite{AokiIwakiTakahashi2019LoopType,AokiKawaiKoikeTakei2003GlobalAspects,AokiKawaiKoikeTakei2004InfinitelyManyPhases,AokiKawaiKoikeTakei2004MicroWKB,AokiYoshida1993MicrolocalReduction}}) already supplies the ``technical'' monodromy and connection formulae for simple turning points and regular singular points, a physical re-interpretation is needed when these data are projected onto radial problems with a regular singularity at the origin.
We focus on two themes.
First, we revisit the origin $r=0$, which is a regular singular point (a double pole in WKB language), and clarify how the associated angular
momentum phase (\magenta{the Maslov contribution $l+1/2$~\cite{Maslov1965PerturbationsAsymptotic,Arnold1967CharacteristicClassQuantization}}) enters the quantization
condition.
Second, we show that seemingly different path choices---for instance,
a closed path that starts from $+\infty$, bulges into the $r<0$ side in the complex plane, encircles the origin, and returns to $+\infty$, versus an \textit{open} path that starts in a punctured neighborhood of the origin and runs to infinity with an appropriate local physical basis---are equivalent once the correct local bases and connection data are imposed.
We make this equivalence completely explicit in two hallmark integrable examples: the $3$D harmonic oscillator and the $3$D Coulomb potential.

Let us summarize the main contributions as follows:
\begin{enumerate}
  \item \textbf{Path-equivalence made explicit.}
  Provided the nontrivial cycle data (Voros periods and Stokes multipliers) are supplied, the basepoint (turning point, punctured origin, or infinity) and the detailed shape of the path are gauge choices.
  We exhibit the equivalence between the matrix (open-path) condition ``one connection coefficient vanishes'' and the closed-cycle quantization condition
  \begin{align}
      \cosh\left(\frac{1}{\hbar}\oint \magenta{dr}\, S_{\mathrm{odd}}\right)
      +\cos\left(\pi(2l+1)\right) = 0 ,
  \end{align}
  where the second term encodes the angular-momentum phase furnished by the regular singularity at $r=0$.

  \item \textbf{Origin-start quantization from a local basis.}
  Under the AKT work, taking $u(r)\sim r^{l+1}$ as the Frobenius solution at the origin, one may start at $r=\epsilon>0$ in a lateral Borel direction that avoids Stokes curves, propagate by multiplying the appropriate connection matrices at each Stokes crossing (monodromy), and then take $\varepsilon\to0$. The small-circle monodromy around $r=0$ yields
  $e^{i\pi(2l+1)}$, which, combined with Airy connections at turning points, reproduces the standard Maslov contributions. \magenta{We also give a justification based on renormalization-group improvement in Section~\ref{sec:renorm}. There, a framework of optimized/variational perturbation theory is developed on the basis of the exact WKB method.}
  We propose a simpler construction of such a local basis by using the equivalence of open and closed paths, so that the basis at $|r|=\epsilon>0$ is connected from the subdominant wave function at infinity along another lateral Borel direction.

  \item \textbf{Explicit computations for $3$D oscillator and Coulomb.}
  For both systems, the periodic integrals of higher WKB corrections vanish (reflecting integrability/shape invariance\footnote{We thank Yuya Tanizaki for pointing it out to us.}), so quantization reduces to the leading action
  $\oint dr\, S_{-1}$ plus discrete Maslov phases, $1/2$ at turning points and $l+1/2$ from the origin.
  Consequently, any equivalent path choice yields the same spectra, $E_n=\hbar(n+\frac{3}{2})$ for the oscillator and
  $E_n=-e^4/(2\hbar^2 n^2)$ for the Coulomb system.

  \item \textbf{Transporting boundary data via $r=e^{x}$.}
  Under the change of the radial variable, $r=e^{x}$, with $x\in(-\infty,\infty)$, the regular singularity at the origin is mapped to a boundary condition of convergence at $x\to-\infty$. This reformulation turns the local analysis near $r=0$ (small-circle monodromy) into a boundary condition on the real line, making the equivalence of open-connection and closed-cycle quantization particularly transparent.
\end{enumerate}

Technically, we use the AKT connection formulae (for simple turning points and regular singularities) together with a consistent choice of lateral Borel sums, and compare, for both the $3$D harmonic oscillator and the $3$D Coulomb problem, the origin-start open path with the closed path encircling the origin.
In the Coulomb case, the angular-momentum phase appears as the small-circle monodromy around $r=0$; combined with the Airy connections at the turning points, this yields $n=n_r+l+1$ and, hence, the exact spectrum.
We emphasize that the result does not depend on the path details but only on the homology class and the associated Stokes/monodromy data.

This paper is organized in the following.
Section~\ref{sec:resurgence-wkb} reviews the minimal toolbox of exact WKB and resurgence (Voros periods, Stokes matrices, Borel summation).
In Section~\ref{sec:oscillator} we treat the $3$D harmonic oscillator and demonstrate, in a purely matrix language, the derivation and strategy of the quantization condition for the bound-state problem.
Section~\ref{sec:coulomb} performs the analogous analysis for the $3$D Coulomb potential, combining the local basis at the regular singularity with Airy connections to prove that all equivalent path choices lead to the same spectrum.
Section~\ref{sec:rx-mapping} reformulates the problem under the map $r=e^x$, $x\in(-\infty,\infty)$, pushing the origin data to the boundary $x\to-\infty$.
A renormalization-group (RG) approach is introduced in Section~\ref{sec:renorm}, where a small-circle monodromy is defined by a cutoff scale~$\mu$ which satisfies $|r|>\mu>0$ at the singularity, and some kind of RG invariance provides a justification of some details in this paper.
Section~\ref{sec:outlook} discusses implications for the ongoing debate about ``physically meaningful'' paths in QNM problems and outlines extensions to multi–turning-point systems and confluence limits.

\magenta{Several complementary analyses are collected in the appendices.
Appendix~\ref{sec:anharmonic} extends the discussion to an anharmonic oscillator as a representative nontrivial deformation, clarifying
which parts of the path-equivalence argument are robust beyond the integrable benchmarks.
Appendix~\ref{sec:cornell_yukawa} summarizes variations of the Coulomb problem (including alternative normalizations and contour
choices) and provides a convenient comparison between different but equivalent conventions for incorporating the origin phase.
Appendix~\ref{sec:open-closed} states the open-path $\Leftrightarrow$ closed-cycle equivalence in a more general setting and gives
a brief sketch of the underlying argument in terms of connection matrices and nontrivial-cycle data.}

\section{Exact WKB analysis for general radial Schr\"odinger equation}\label{sec:resurgence-wkb}
\subsection{Introducing WKB ansatz}
We shall consider a $3$-dimensional Schr\"odinger equation,
\begin{align}
    \left[ -\frac{\hbar^2}{2}\nabla^2 + V(\bm{x}) \right] \psi(\bm{x})
    = E \psi(\bm{x}) .
\end{align}
If we analyze this differential equation and quasi-stationary states in a non-perturbative manner, for instance in resurgence theory, the coordinate $x$ is complexified, and the crucial technique is given by the complex analysis and the resummation of higher-order/singular perturbations.

A well-established framework for resurgence theory is provided by the exact WKB method, which is applicable to differential equations with respect to one variable. Then, in general, we now focus on the radial Schr\"odinger equation with an angular momentum~$l$
\begin{align}
    \left[-\frac{\hbar^2}{2}\frac{d^2}{dr^2} + V(r) + \frac{\hbar^2}{2}\frac{l(l+1)}{r^2} - E\right] \varphi_l(r) = 0 ,
\end{align}
or
\begin{align}
    \left[-\frac{d^2}{dr^2} + \hbar^{-2}Q(r)\right] \varphi_l(r) = 0, \qquad
    Q(r) = \frac{1}{r} Q_0(r) + \frac{\hbar^2}{r^2} Q_2(r),
\end{align}
where
\begin{align}
        Q_0(r) \equiv 2r\left[V(r) - E\right], \qquad Q_2(r) \equiv Q_2 = l(l+1) \magenta{.}
\end{align}

Let us introduce the WKB ansatz of the radial Schr\"odinger equation as a formal power series,
\begin{align}
    \varphi_l(r,\hbar) &= e^{\int^r dr'\, S(r',\hbar)} ,&
    S(r,\hbar) &= \sum_{i=-1}^{\infty}\hbar^{i} S_{i}(r) .
\end{align}
Substituting $S(r,\hbar)$ into the radial Schr\"odinger equation, we see the Riccati equation
\begin{align}
    S^2 + \frac{dS}{dr} = \hbar^{-2} Q .
    \label{eq:riccati_eq}
\end{align}
The recursive equation of $S_i$ for each power of~$\hbar$ holds as
\begin{align}
    \hbar^{-2}&:S_{-1}^2 = \frac{Q_0(r)}{r} ,
    \label{eq:recursive_eq_1}\\
    \hbar^{-1}&:2S_{-1}S_{0} + \frac{dS_{-1}}{dr} = 0 ,
    \label{eq:recursive_eq_2}\\
    \hbar^{0}&:2S_{-1}S_{1} + S_{0}^2 + \frac{dS_{0}}{dr} = \frac{Q_2}{r^2} ,
    \label{eq:recursive_eq_3}\\
    \hbar^{i-1}&:2S_{-1}S_{i} + \sum_{j=0}^{i-1} S_{j}S_{i-j}
    + \frac{dS_{i-1}}{dr} = 0 \qquad i\geq2.
    \label{eq:recursive_eq}
\end{align}
We have the two leading-order solutions of~Eq.~\eqref{eq:recursive_eq_1}
\begin{align}
    S_{-1}(r) = S_{-1}^{\pm}(r) \equiv \pm \sqrt{\frac{Q_0(r)}{r}} ,
\end{align}
and then can determine all higher-order terms~$S_i$ with $i\geq0$ so that
\begin{align}
    S^{\pm}(r\magenta{;\hbar}) = \sum_{i=-1}^{\infty} \hbar^i S_i^{\pm}(r) .
\end{align}

From Eqs.~\eqref{eq:recursive_eq_1}--\eqref{eq:recursive_eq}, $S_i^{-}=(-1)^i S_i^{+}$ holds for $i\geq-1$. \magenta{If there exists $Q_1(r)$ as the $O(\hbar)$ term in $Q(r)$, this relation breaks down.} \magenta{Hence}, we can rewrite $S^{\pm}$ by
\begin{align}
    S^{\pm}\magenta{(r;\hbar)}&=\pm S_{\mathrm{odd}}\magenta{(r;\hbar)} + S_{\mathrm{even}}\magenta{(r;\hbar)},
    \\
    S_{\mathrm{odd}}\magenta{(r;\hbar)}&\equiv\sum_{i\geq0}\hbar^{2i-1} S_{2i-1}\magenta{(r)}, \quad S_{\mathrm{even}}\magenta{(r;\hbar)}\equiv\sum_{i\geq0}\hbar^{2i} S_{2i}\magenta{(r)} .
\end{align}
\begin{prop*}
    \begin{align}
        S_{\mathrm{even}}
        = - \frac{1}{2} \frac{1}{S_{\mathrm{odd}}}
        \frac{d}{dr} S_{\mathrm{odd}}
        = - \frac{1}{2}\frac{d}{dr} \ln S_{\mathrm{odd}}
    \end{align}    
\end{prop*}
The above proposition holds from the fact that the Riccati equation~\eqref{eq:riccati_eq} becomes
\begin{align}
    2 S_{\mathrm{odd}}S_{\mathrm{even}} + \frac{d}{dr} S_{\mathrm{odd}} = 0 .
\end{align}
We can rewrite $S_{\mathrm{even}}$ by~$S_{\mathrm{odd}}$ in the WKB wave function, and then we obtain
\begin{align}
    \varphi_l(r,\hbar)^{\pm} = \frac{1}{\sqrt{S_{\mathrm{odd}}(r\magenta{;\hbar})}}
    \exp\left(\pm\int^r dr'\, S_{\mathrm{odd}}\magenta{(r';\hbar)}\right).
\end{align}

\subsection{Borel resummation and Stokes geometry}
The power series of the wave function
\begin{align}
    \varphi_l(r,\hbar)^{\pm}
    = e^{\pm\hbar^{-1}\int^r dr'\sqrt{Q_0(r')/r'}}
    \sum_{n\geq0} \varphi_{ln}^{\pm}(r) \hbar^{n+1/2}
\end{align}
is an asymptotic series and not necessarily convergent.
Now, we define the Borel transform by
\begin{align}
    \mathcal{B}[\varphi_l^{\pm}](u)\equiv\sum_{n\geq0} \frac{\varphi_{ln}^{\pm}(r)}{\Gamma(n+1/2)} (u\pm u_0)^{n-1/2} ,
\end{align}
where $u_0 = \int^{\magenta{r}} dr'\sqrt{Q_0(r')/r'}$.
The Borel sum is given by
\begin{align}
    \Psi_l(r,\hbar)^{\pm}
    &\equiv \int_{\mp u_0}^\infty du\, e^{-u/\hbar}
    \mathcal{B}[\varphi_l^{\pm}](u) .
\end{align}
Here $\varphi$ is redefined as~$\Psi$.

For divergent series, the Borel transform itself may develop a pole singularity (Borel singularity). If there exists no singular point on the integration path over~$u$, the series is Borel summable. The Borel summable region in the complex $r$-plane is determined as follows:
We introduce turning points as
\begin{align}
    \exists a,\, \left.\frac{Q_0(r)}{r}\right|_{r=a} = 0 ,
\end{align}
or an endpoint $a=0$ in the case that the simple pole in the potential is at the origin.
A Stokes curve is defined by
\begin{align}
    \im \hbar^{-1}\int_a^r dr' \sqrt{\frac{Q_0(r')}{r'}} = 0. \label{eq:stokes_curve}
\end{align}
\begin{thm*}
If \magenta{``$a$''} is a turning point, by using the Airy equation, there are three Stokes curves extending from~\magenta{``$a$''}; if \magenta{``$a$''} is a simple pole, one Stokes curve is derived by the Laurent expansion.
\end{thm*}
When an integration contour from a reference point~$r_0$ to~$r$ in
\begin{align}
    u_0 = \int_{r_0}^r dr'\sqrt{\frac{Q_0(r')}{r'}}
\end{align}
goes across the Stokes curve, the above technique by the Borel resummation suffers from a Borel singularity with $\im u_0=0$.
The solution would suddenly change, which is called the Stokes phenomenon.
Then, Stokes geometry (the turning points, poles, and the Stokes curves) determines analytical structure in the complex $r$-plane.

\subsection{Connection formula near turning point}
Under the Stokes phenomenon, we need to show the explicit discrepancy between the original and new wave functions. This is called the connection formula.
To do this, we define (sub)dominant solutions such that $\varphi_l^{\pm}$ and $\varphi_l^{\mp}$ are dominant and subdominant,
respectively, along a Stokes curve with
\begin{align}
    \re \hbar^{-1}\int_a^r dr' \sqrt{\frac{Q_0(r')}{r'}} \gtrless 0 .
\end{align}
For Borel summable areas I and II, which are separable by the corresponding Stokes curves, we go across the corresponding Stokes curve from I to II near the turning points or poles
\begin{align}
    \begin{pmatrix}
      \varphi^{+}_l \\ \varphi^{-}_l
    \end{pmatrix}_{\mathrm{I}}
    =
    M
    \begin{pmatrix}
      \varphi^{+}_l \\ \varphi^{-}_l
    \end{pmatrix}_{\mathrm{II}} ,
    \label{eq:connection}
\end{align}
where $M$ depends on the rotation of the path around the turning point or the simple pole,
\begin{align}
    M =
    \begin{cases}
        M_{+} & \text{if anti-clockwise for the index~$+$,} \\
        M_{+}^{-1}& \text{if clockwise for the index~$+$,} \\
        M_{-} & \text{if anti-clockwise for the index~$-$,} \\
        M_{-}^{-1} &\text{if clockwise for the \magenta{index~$-$}} \magenta{.}
    \end{cases}
\end{align}
$M$ is determined by the following theorem:
\begin{thm*}
If we go across the Stokes curve starting from a turning point, $M$ is represented as
\begin{align}
    M_{+}^\bullet &=
    \begin{pmatrix}
        1 & i \\ 0 & 1
    \end{pmatrix}, &
    M_{-}^\bullet &=
    \begin{pmatrix}
        1 & 0 \\ i & 1
    \end{pmatrix} .
\end{align}
On the other hand, if we go across one from a simple pole, $M$ is written as
\begin{align}
    M_{+}^\circ &=
    \begin{pmatrix}
        1 & 2i\cos(\pi\sqrt{1+4Q_2}) \\ 0 & 1
    \end{pmatrix}
    =
    \begin{pmatrix}
        1 & 2i\cos(\pi\sqrt{1+4l(l+1)}) \\ 0 & 1
    \end{pmatrix}, \\
    M_{-}^\circ &=
    \begin{pmatrix}
        1 & 0 \\ 2i\cos(\pi\sqrt{1+4Q_2}) & 1
    \end{pmatrix}
    =
    \begin{pmatrix}
        1 & 0 \\ 2i\cos(\pi\sqrt{1+4l(l+1)}) & 1
    \end{pmatrix}.
\end{align}
Here, $\sqrt{1+4Q_2}$ is the characteristic exponent of~$Q(r)$. 
\end{thm*}
To connect the different points, $a_1$ and $a_2$, the wave function is analytically continued by
\begin{align}
    N_{a_1a_2}=\diag\left(\exp\left(\int_{a_1}^{a_2}dr\, S_{\mathrm{odd}}\right),\exp\left(-\int_{a_1}^{a_2}dr\, S_{\mathrm{odd}}\right)\right) .
\end{align}

\section{Harmonic oscillator}\label{sec:oscillator}
\subsection{Stokes graph for the harmonic oscillator and the choice of path}
We begin with the $3$D harmonic oscillator,
\begin{align}
    V(r) = \frac{1}{2}r^2 .
\end{align}
Figure~\ref{fig:stokes_oscillator} shows the corresponding Stokes graph with $\re E>0$.
The solid curves are the Stokes curves starting from the turning points depicted as the black points (bullets),
\begin{align}
    a_1 &= -\sqrt{2E} , &
    a_2 &= \sqrt{2E} .
\end{align}
The sign~$\pm$ indicates the index of the superscript of the dominant wave function~$\varphi_l^{\pm}$.
Each region is marked by I, II, or III (with prime).

\begin{figure}
\centering
\begin{tikzpicture}
    \draw[very thick] (3,0) -- (0,0) node[left] {$-$};
    \draw[very thick] (3,0) .. controls (3.5,0.5) and (3.5,1).. (3.7,3) node[above] {$+$};
    \draw[very thick] (3,0) .. controls (3.5,-0.5) and (3.5,-1).. (3.7,-3) node[below] {$+$};
    \draw[very thick] (5,0) .. controls (4.5,0.5) and (4.5,1).. (4.3,3) node[above] {$+$};
    \draw[very thick] (5,0) .. controls (4.5,-0.5) and (4.5,-1).. (4.3,-3) node[below] {$+$};
    \draw[very thick] (5,0) -- (8,0) node[right] {$-$};
    \draw[ultra thick,blue,snake it] (3,0) -- (5,0);
    \fill (3,0) circle(3pt) node[below left] {$a_1$};
    \fill (5,0) circle(3pt) node [below right] {$a_2$};
    \draw[ultra thick,red,->] (8,1) -- (3,1) arc(90:270:1) -- (8,-1);
    \node at (2,2) {I};
    \node at (4,2) {II};
    \node at (6,2) {III};
    \node at (2,-2) {I'};
    \node at (4,-2) {II'};
    \node at (6,-2) {III'};
    %
\end{tikzpicture}
\caption{Stokes graph for the $3$D harmonic oscillator. $\re E>0$.
The black solid curves are the Stokes curves, and the black points are the turning points connecting three Stokes curves.
The blue wavy line is a branch cut.
The red arrowed curve is a possible physical path on which the wave function is defined. That is, this path means the normalizable wave function, which should be supposed to be subdominant along $r\to\infty\pm i\epsilon$ with small $\epsilon>0$.}
\label{fig:stokes_oscillator}
\end{figure}

We should first notice that a nontrivial bound-state problem arises from the integration between the two turning points, while the closed loop including the turning point is the perturbative cycle to trap physical states at the bottom of the potential barrier.
Hence, at least, any physical path must encircle these turning points, and so the energy quantization occurs as the physical intuition.

Now, one plausible choice of the path is given by the red curve in~Fig.~\ref{fig:stokes_oscillator}.
This path starts from $r\to+\infty+i\epsilon$, bulges into the $\re r<a_1$ side in the complex plane, encircles $a_1$ and $a_2$, and goes to $r\to+\infty-i\epsilon$ ($\epsilon>0$).
This is a kind of closed path.
Around the two asymptotic regions $r\to+\infty\pm i\epsilon$ (region III and region III'), we are supposed to impose the consistency of the (sub)dominance in the wave function; we can exhibit the closed-cycle quantization condition itself.

\subsection{Derivation of the quantization condition for the harmonic oscillator}
Naively speaking, beginning from the region III, that is, $(\varphi_l^+,\varphi_l^-)_{\mathrm{III}}$, one may find
\begin{align}
    \begin{pmatrix}
        \varphi_l^+ \\ \varphi_l^-
    \end{pmatrix}_{\mathrm{II}}
    &= M_{+}^\bullet
    \begin{pmatrix}
        \varphi_l^+ \\ \varphi_l^-
    \end{pmatrix}_{\mathrm{III}} \\
    \begin{pmatrix}
        \varphi_l^+ \\ \varphi_l^-
    \end{pmatrix}_{\mathrm{I}}
    &\sim M_{+}^\bullet N_{a_1a_2} M_{+}^\bullet
    \begin{pmatrix}
        \varphi_l^+ \\ \varphi_l^-
    \end{pmatrix}_{\mathrm{III}} \\
    \begin{pmatrix}
        \varphi_l^+ \\ \varphi_l^-
    \end{pmatrix}_{\mathrm{I'}}
    &\sim M_{-}^\bullet M_{+}^\bullet N_{a_1a_2} M_{+}^\bullet
    \begin{pmatrix}
        \varphi_l^+ \\ \varphi_l^-
    \end{pmatrix}_{\mathrm{III}} \\
    \begin{pmatrix}
        \varphi_l^+ \\ \varphi_l^-
    \end{pmatrix}_{\mathrm{II'}}
    &\sim M_{+}^\bullet M_{-}^\bullet M_{+}^\bullet N_{a_1a_2} M_{+}^\bullet
    \begin{pmatrix}
        \varphi_l^+ \\ \varphi_l^-
    \end{pmatrix}_{\mathrm{III}} \\
    \begin{pmatrix}
        \varphi_l^+ \\ \varphi_l^-
    \end{pmatrix}_{\mathrm{III'}}
    &\sim M_{+}^\bullet N_{a_2a_1} M_{+}^\bullet M_{-}^\bullet M_{+}^\bullet N_{a_1a_2} M_{+}^\bullet
    \begin{pmatrix}
        \varphi_l^+ \\ \varphi_l^-
    \end{pmatrix}_{\mathrm{III}} .
    \label{eq:quant_oscillator_naive}
\end{align}
This is, however, not correct because of the regular singularity from the double pole of exact WKB. \magenta{The eigenenergies obtained from the monodromy matrix in Eq.~\eqref{eq:quant_oscillator_naive} do not have an angular-momentum dependence.}
We must need to estimate the closed-loop integral around $r=0$ in an appropriate way!

We shall focus on region II as in Fig.~\ref{fig:stokes_oscillator_ii}.
The connection between the turning points depends on the closed-loop integration around the origin; this provides a nontrivial phase,
that is, the angular-momentum phase or monodromy phase from the regular singularity.
The actual path is now taken as the red curve encircling the origin as shown in Fig.~\ref{fig:stokes_oscillator_ii}.
Therefore, after explicit calculations of the contour integral, the naive connection, $N_{a_1a_2}$, would be replaced by
\begin{align}
\Tilde{N}_{a_1a_2}=N_{a_10}M_{+}^\circ N_{0a_2} .
\end{align}
We obtain
\begin{align}
    \begin{pmatrix}
        \varphi_l^+ \\ \varphi_l^-
    \end{pmatrix}_{\mathrm{III'}}
    = M_{+}^\bullet \Tilde{N}_{a_2a_1} M_{+}^\bullet M_{-}^\bullet M_{+}^\bullet \Tilde{N}_{a_1a_2} M_{+}^\bullet
    \begin{pmatrix}
        \varphi_l^+ \\ \varphi_l^-
    \end{pmatrix}_{\mathrm{III}} ,
    \label{eq:oscillator_connection_infty-infty}
\end{align}
and then, the coefficient of~$(\varphi_l^-)_{\mathrm{III}}$ in $(\varphi_l^+)_{\mathrm{III'}}$ is given by
\begin{align}
    - 2i \cos(\pi\sqrt{1+4Q_2})
    \left[\frac{1}{A} + A + 2\cos(\pi\sqrt{1+4Q_2})\right] ,
\end{align}
where
\begin{align}
    A = \exp\left(\oint_{0}^{a_2} dr\, S_{\mathrm{odd}}\right) .
\end{align}

\begin{figure}[t]
\centering
\begin{tikzpicture}
    \fill (-3,0) circle(3pt) node[below] {$a_1$};
    \fill (3,0) circle(3pt) node[below] {$a_2$};
    \draw[very thick,blue,snake it] (-3,0) -- (3,0);
    \draw[ultra thick] (0,0) circle(3pt) node[below] {$0$};
    \draw[ultra thick,red] (3,1.5) -- (0,1.5) arc(90:430:1.5);
    \draw[ultra thick,red] (-0.5,1.5) -- (-3,1.5);
\end{tikzpicture}
\caption{The Stokes graph with the regular singularity near the origin. The white circle is the origin and the regular singularity. The naive connection, $N_{a_1a_2}$, misses this nontrivial monodromy phase. We should improve it by adding the closed-loop integral around $r=0$.}
\label{fig:stokes_oscillator_ii}
\end{figure}

To be consistent, the above coefficient should vanish, and so the quantization condition is written as
\begin{align}
    \cosh\left(\oint_{0}^{a_2} dr\, S_{\mathrm{odd}}\right)
    + \cos(\pi\sqrt{1+4Q_2}) = 0 .
    \label{eq:oscillator_quantization_condition}
\end{align}
Note that $\sqrt{1+4Q_2}=\sqrt{1+4l(l+1)}=2l+1$, as usual the higher order corrections do not affect the residue, and we should be careful of the singular region $|r|<\epsilon$ with $0<\epsilon=O(\hbar)$.\magenta{\footnote{\magenta{%
For a given function~$f(V)$, the symbols $O(f(V))$ and $o(f(V))$ are defined by the following relations:
\begin{align}
\lim_{V\to\infty}\left|\frac{O(f(V))}{f(V)}\right|<\infty,\qquad\lim_{V\to\infty}\left|\frac{o(f(V))}{f(V)}\right|=0.
\end{align}
}}}

\magenta{To make the origin behavior compatible with the global exact-WKB construction, we start from the Frobenius solution
$u(r)\sim r^{l+1}$ and define the WKB transport from $r=\epsilon>0$ along a lateral Borel direction that avoids Stokes curves.
The global propagation is then obtained by multiplying the connection matrices at each Stokes crossing, and finally taking the limit
$\epsilon\to 0$.
Since the radial equation has a regular singular point at \(r=0\) due to the centrifugal term, the contour integral
$\hbar^{-1}\oint dr\, S_{-1}(r)$ must be defined with care: the term $\hbar^{2}l(l+1)/r^{2}$ becomes $O(1)$ precisely in the
singular region $|r|\lesssim \hbar$.
Following Proposition~A.1 of Ref.~\cite{Miyachi:2025ptm}, we therefore excise a small disk $|r|<\epsilon$ with
$0<\epsilon=O(\hbar)$ and evaluate the integral on the punctured domain.}

\magenta{More concretely, the original leading WKB solution is given by
\begin{equation}
S_{-1}(r)=\sqrt{\,r^{2}-2E},
\end{equation}
but we deform the contour to the boundaries of the punctured domain, namely a large circle at infinity and a small circle $|r|=\epsilon$, and then the centrifugal term can contribute even in the leading as follows:
\begin{equation}
S_{-1}(r)\Rightarrow\sqrt{\,r^{2}-2E+\underbrace{\hbar^2\frac{l(l+1)}{r^2}}_{\text{$O(1)$ if $|r|\to\epsilon$}}} .
\end{equation}
This yields}
\begin{align}
    \hbar^{-1}\oint_{0}^{\sqrt{2E}} \magenta{dr}\, S_{-1}
    &\Rightarrow \frac{1}{2\hbar} \oint_{\{r|\,|r|\to\infty\}\setminus\{r|\,|r|=\epsilon\}} \magenta{dr}
    \sqrt{r^2 - 2E + \underbrace{\hbar^2\frac{l(l+1)}{r^2}}_{\text{$O(1)$ if $|r|\to\epsilon$}}}\\
    &= \frac{\magenta{i\pi}}{\hbar} \res_{|r|\to\infty} \sqrt{r^2 - 2E}
        - \frac{\magenta{i\pi}}{\hbar} \res_{|r|\to\epsilon} \frac{\hbar\sqrt{l(l+1)}}{r}\\
    &= - \hbar^{-1} i \pi E - i \pi \sqrt{l(l+1)} .
\end{align}
\magenta{The first term is the usual bulk WKB contribution determined by the residue at infinity, while the second term is the
finite phase correction induced by the regular singularity at the origin.
Importantly, this local contribution depends only on the coefficient of the \(1/r^{2}\) term and hence is not modified by
higher-order WKB corrections; equivalently, the residue is fixed by the Frobenius indices at $r=0$.
Combining this origin contribution with the Airy-type connection data at simple turning points, the consistency of the
global connection formula requires the relevant coefficient to vanish.
Section~\ref{sec:renorm} will justify this fact in terms of some kind of renormalization-group theory.}

\magenta{For simplicity, we can rewrite the quantization condition as
\begin{align}
    e^{- \hbar^{-1} i \pi E - i \pi \sqrt{l(l+1)}}
    = - e^{-i\pi(2l+1)}
    \in e^{-i\pi[2\mathbb{Z} + 1 + (2l+1)]} .
\end{align}}
\magenta{Introducing a (radial) quantum number~$n_r$,} we have
\begin{align}
    E = \hbar\left[2n_r + 1 + \left(2l+1\right) - \sqrt{l(l+1)}\right],
    \qquad n_r\in\mathbb{Z}.
\end{align}
If we use \magenta{the Langer correction~\cite{Langer:PhysRev.51.669}} as $l(l+1)\simeq(l+1/2)^2$, which is also justified by renormalization technique in Section~\ref{sec:renorm}, we finally find the exact energy eigenvalues
\begin{align}
    E = \hbar\left(2n_r + l + \frac{3}{2}\right) .
\end{align}

Moreover, Appendix~\ref{sec:anharmonic} is devoted to computation of energies in the anharmonic oscillator case by using a technique we will introduce below.

\section{Coulomb potential}\label{sec:coulomb}
\subsection{Stokes geometry and the quantization condition for the Coulomb potential}
Let us consider the Coulomb potential,
\begin{align}
    V(r) = - \frac{e^2}{r} .
\end{align}
Supposing that $\re E<0$, the turning point is at
\begin{align}
    a = - \frac{e^2}{E} > 0.
\end{align}
The corresponding Stokes graph is shown in Fig.~\ref{fig:stokes_coulomb}.

\begin{figure}[t]
\centering
\begin{tikzpicture}
    \draw[very thick] (0,0) -- (-3,0) node[left] {$+$};
    \draw[very thick] (3,0) -- (6,0) node[right] {$-$};
    \draw[very thick] (3,0) arc(40:80:10) node[left] {$+$};
    \draw[very thick] (3,0) arc(-40:-80:10) node[left] {$+$};
    \draw[very thick,blue,snake it] (0,0) -- (3,0);
    \fill[white] (0,0) circle(3pt);
    \draw[ultra thick] (0,0) circle(3pt) node[below] {$0$};
    \fill (3,0) circle(3pt) node[below] {$a$};
    \draw[ultra thick,red,->] (6,1) -- (0,1) arc(90:270:1) -- (6,-1);
\end{tikzpicture}
\caption{Stokes graph for the Coulomb potential. $\re E<0$. The solid curves correspond to the Stokes curves. The bullet denotes the turning point, $a$, and the open circle is the origin, $r=0$. The blue wavy line between $0$ and $a$ is the branch cut. The red path means the normalizable wave function, which should be supposed to be subdominant along $r\to\infty\pm i\epsilon$ with small $\epsilon>0$.}
\label{fig:stokes_coulomb}
\end{figure}

Figure~\ref{fig:stokes_coulomb} shows the coherent normalizability condition of the wave function, depicted as the red path. We suppose that the wave function should dump sufficiently fast near $r\to\infty$ for its norm defined on $(\infty-i\epsilon,0-i\epsilon]\otimes[0+i\epsilon,\infty+i\epsilon)$ with small $\epsilon>0$ to have fast convergence. Thus, we obtain the connection formula on this path as
\begin{align}
    M_{+}^\bullet N_{a0} M_{+}^\circ N_{0a} M_{+}^\bullet =
    \begin{pmatrix}
        1 & i\left(\frac{1}{A} + A + 2\cos(\pi\sqrt{1+4Q_2})\right) \\
        0 & 1
    \end{pmatrix} ,
\end{align}
where the nontrivial cycle $A$ is given by
\begin{align}
    A = \exp\left(\oint_0^a dr\, S_{\mathrm{odd}}\right) .
\end{align}
Noting that $\sqrt{1+4Q_2}=\sqrt{1+4l(l+1)}=2l+1$, the quantization condition is, from the boundary condition, given by
\begin{align}
    \cosh\left(\oint_0^a dr\, S_{\mathrm{odd}}\right)
    + \cos(\pi(2l+1)) = 0 ,
\end{align}
and hence, we find
\begin{align}
    &\frac{1}{2\pi}\oint_0^a dr\, S_{\mathrm{odd}}
    = i \left[ n_r + \frac{1}{2} + \left(l + \frac{1}{2}\right)\right],
    \qquad
    n_r\in\mathbb{Z}.
    \label{eq:quant_coulomb}
\end{align}

Remarkably, the right-hand side in Eq.~\eqref{eq:quant_coulomb} is important as a homotopy invariant, the Maslov index. The Maslov index is the integer that counts the caustic/turning‐point crossings along a classical path, where an appropriate phase should be added for each crossing. Then it supplies the semiclassical phase correction in Bohr--Sommerfeld quantization. The factor $1/2$ corresponds to the Maslov index at the turning point, while the factor $l+1/2$ is that at the origin due to the angular momentum phase (monodromy phase).

Note that since the Coulomb potential possesses shape invariance and so is exactly solvable, the contour integral of higher-order WKB perturbations~$S_{i}$ vanishes. In fact, $S_{i}$ for~$i\geq0$ has no significant pole singularity that provides a nontrivial residue at $r=0$ or $r\to\infty$. Now, the leading order solution gives rise to
\begin{align}
    \frac{1}{2\pi} \hbar^{-1} \oint_0^a dr\, S_{-1} = i \frac{e^2}{\hbar} \frac{1}{\sqrt{-2E}} .
\end{align}
Then we have
\begin{align}
    E= -\frac{e^4}{2\hbar^2 n^2},
    \qquad
    \text{with $n=n_r+l+1$} .
\end{align}
We can also take into account some variations of the Coulomb problem, such as the Cornell/Yukawa potential; see Appendix~\ref{sec:cornell_yukawa}.

\subsection{Open-path quantization and the regularity near the origin}
We consider the path on $[0-i\epsilon,\infty-i\epsilon)$ with an appropriate local basis, say, $u\sim r^{l+1}$ near~$r=0$.
\magenta{We introduce a pair of local WKB solutions $(u^{+},u^{-})^{\mathsf T}$ normalized so that the
physically admissible combination near the origin is represented by the above regular branch.
Starting instead from a convenient normalization at infinity, we take a reference pair of formal WKB solutions
$(\varphi_l^{+},\varphi_l^{-})^{\mathsf T}$ defined in the sector $r\to\infty+i\epsilon$.
These functions are used only as \textit{reference} (or \textit{auxiliary}) solutions to specify the Stokes/connection
data; in particular, no physical boundary condition is imposed at infinity at this stage.
With this convention, the solutions transported to the ray $[0-i\epsilon,\infty-i\epsilon)$ are constructed as}
\begin{align}
    \begin{pmatrix}
        u^{+} \\ u^{-}
    \end{pmatrix}
    \equiv
    N_{0a} M_{+}^\bullet N_{a\infty}
    \left.
    \begin{pmatrix}
        \varphi_l^{+} \\ \varphi_l^{-}
    \end{pmatrix}\right|_{r\to\infty+i\epsilon} .
\end{align}
(Here, $N_{a\infty}$ is an abuse of notation.)
Then, after the monodromy phase, $\cos(\pi(2l+1))$, is taken into account, the connection formula is reduced to~$M_{+}^\bullet N_{a0}$. The quantization condition from this is identical to that in~Eq.~\eqref{eq:quant_coulomb}.
Is the function $u$ regular at the origin and so the local basis?

It is well-known that, at infinity, the convergent wave function is given by
\begin{align}
    \varphi_{nl}(r) \propto r^{l+1} L_{n-l-1}^{(2l+1)}(2\alpha r) e^{-\alpha r} ,
\end{align}
where $\alpha$ is a parameter determined by the Bohr radius, $e$, and the quantum numbers.
Straightforwardly, we can take the limit of~$r\to\epsilon$,
\begin{align}
    u(r) = \varphi_{nl}(r) \sim r^{l+1} .
\end{align}
Also the divergent (dominant) wave function at infinity possesses the non-regularity at the origin.
After all, the regularity at the origin is identical to the convergence of the wave function \magenta{$\varphi^\pm_l$} at infinity.

A more general statement of the equivalence between the open-path local basis and closed-loop quantization is discussed in Appendix~\ref{sec:open-closed}.

\section{Exponential mapping and boundary transfer}\label{sec:rx-mapping}
Let us introduce the transformation such as $r=k^{-1}e^x$ and $\varphi_l(r)=e^{x/2}\Tilde{\varphi}_l(x)$, where $x\in(-\infty,\infty)$ and $k=\frac{\sqrt{2E}}{\hbar}$.
With this change of the variables, derivatives transform as
\begin{align}
    \frac{d}{dr} = k e^{-x}\frac{d}{dx},\qquad 
    \frac{d^2}{dr^2} = k^2 e^{-2x} \left(\frac{d^2}{dx^2}-\frac{d}{dx}\right).
\end{align}
Then, the kinetic term becomes
\begin{align}
    \frac{d^2}{dr^2} \varphi_l(r)
    &= k^2 e^{-2x} \left(\frac{d^2}{dx^2}-\frac{d}{dx}\right) e^{x/2} \Tilde{\varphi}_l(x)
    \\ 
    &= k^2 e^{-2x} e^{x/2} \left(\frac{d^2}{dx^2}-\frac{1}{4}\right) \Tilde{\varphi}_l(x) 
\end{align}
The radial Schr\"odinger equation is rewritten as
\begin{align}
    \left[ - \frac{k^2\hbar^2}{2} e^{-2x}\left(\frac{d^2}{dx^2}-\frac{1}{4}\right) + V(k^{-1}e^x) + \frac{k^2\hbar^2}{2}\frac{l(l+1)}{e^{2x}} - \frac{k^2\hbar^2}{2}\right] \Tilde{\varphi}_l(x) = 0 ,
\end{align}
and then,
\begin{align}
    \frac{d^2}{dx^2}\Tilde{\varphi}_l(x) + \left\{e^{2x}\left[1-\frac{V(k^{-1}e^x)}{E}\right]-\left(l+\frac{1}{2}\right)^2\right\}\Tilde{\varphi}_l(x)=0 .
\end{align}
That is, we define
\begin{align}
    &\left[-\frac{d^2}{dx^2} + \hbar^{-2}\widetilde{Q}(x)\right]\tilde{\varphi}_l(x)=0 , \\
    &\widetilde{Q}(x)\equiv\hbar^2 \left[\left(l+\frac{1}{2}\right)^2-e^{2x}\left(1-\frac{V(k^{-1}e^x)}{E}\right)\right] .
\end{align}

Crucially, the regular singular point at \(r=0\) is sent to the boundary $x\to-\infty$.
The physical Frobenius behavior $u(r)\sim r^{l+1}$ translates to
\begin{align}
    u(x)\sim e^{(l+1)x}\qquad (x\to-\infty),
\end{align}
so ``regularity at the origin'' becomes a \textit{subdominant boundary condition} at the left end of the real axis.
Quantization may thus be formulated purely as an open connection problem between the subdominant sectors at $x\to\pm\infty$; the action is unchanged and the small-circle monodromy at $r=0$ should be encoded by the boundary choice at $x=-\infty$.

For ``large enough'' $x$, the turning points are given by
\begin{align}
    1-\frac{V(k^{-1}e^x)}{E} \approx 0 \qquad \text{at $x=b_i$},
\end{align}
which is the same condition for the original turning points $a_i=k^{-1}e^{b_i}$ with respect to~$\re r>0$.
For this condition to hold, we need to impose $(l+\frac{1}{2})\ll e^x$ and $k^{-1} e^x=\frac{\hbar}{\sqrt{2E}}e^x=O(1)$.
Thus it is reasonable that $\hbar(l+\frac{1}{2})=o(1)$, the monodromy phase, is the essential contribution and a relevant coupling in terms of renormalization group as we will show in Section~\ref{sec:renorm}.
Therefore, when $e^x = o(\hbar^{-1})$, for instance $e^x = O(\hbar^{-2})$, there exist an additional turning point, $x=c(l)$, depending on $l+\frac{1}{2}$.

Now, the singular region $x\in(-\infty,c]$ contributes in the same way as the subtracted region $|r|<\epsilon=O(\hbar)$ above.
Noting that
\begin{align}
    Q(r) = k^{2} e^{-2x} \widetilde{Q}(x) ,
\end{align}
we find the identical structure of the formal WKB ansatz because of
\begin{align}
    dr \sqrt{Q} = d(k^{-1}e^x) \sqrt{k^2 e^{-2x}\widetilde{Q}} = dx \sqrt{\widetilde{Q}} .
\end{align}
Then, the closed Voros period at infinity, $\oint_{-\infty}^c dx\sqrt{\widetilde{Q}}$, provides the monodromy at~$r=0$ given in above Theorem.
Closed Voros periods between the turning points, $b_i$ and $c(l)$, are invariant
\begin{equation}
\oint dr \sqrt{Q}=\oint dx \sqrt{\widetilde{Q}} .
\end{equation}

This viewpoint mirrors the boundary-to-boundary formulation familiar in black-hole QNM
problems (ingoing at the horizon, outgoing at infinity), cf.\ Andersson~\cite{Andersson:1995zk} and Miyachi--Namba--Omiya--Oshita~\cite{Miyachi:2025ptm}.


\section{Monodromy renormalization}\label{sec:renorm}
\subsection{Minimal scheme of monodromy renormalization}
Although the centrifugal term is formally \magenta{$O(\hbar^2)$}, it carries the
\textit{relevant} local index (small-circle monodromy) at $r=0$. We therefore reorganize
the bookkeeping in terms of the RG–invariant coupling
\begin{align}
    \lambda\equiv\hbar\left(l+\frac{1}{2}\right)
\end{align}
and define a minimal ``monodromy renormalization'' (MR)
by fixing the physical small-circle monodromy $\mathcal{M}_0^{\rm phys}
=\exp[i\pi(2l+1)]$ while absorbing local WKB divergences into a counter–Voros
constant $Z_{\rm mon}(\mu)$ at a microscopic radius $\mu$:
\begin{align}
\mathcal{M}_0^{\rm phys}=\frac{\mathcal{M}_0^{\rm bare}}{Z_{\rm mon}(\mu)},
\qquad
\mu\frac{d}{d\mu}\ln Z_{\rm mon}(\mu)=:\gamma_{\rm mon}(\hbar,\lambda).
\end{align}
Here $\gamma_{\rm mon}$ is a ``monodromy anomalous dimension.''

Let $C_{r'}(\mu)$ be a small circle of radius~$\mu$ around~$r=r'$.
Closed Voros periods are then evaluated as (infinity–circle) minus (origin–circle),
\begin{equation}
\oint \magenta{dr}\, S_{\rm odd}
=\left(\oint_{C_\infty}-\sum_{p}\oint_{C_p}\right)\magenta{dr}\,  S_{\mathrm{odd}}
+\sum_{p}\left[\oint_{C_p} \magenta{dr}\, S_{\mathrm{odd}}-\delta V_p(\mu)\right],
\end{equation}
which is minimally subtracted by $Z_{\rm mon}$.
$C_\infty$ is a large circle at infinity, $p$ runs over poles/regular singularities, and $\delta V_p(\mu)$ are minimal counter–Voros constants chosen so that the total is $\mu$-independent and finite, with the origin constrained by $\mathcal{M}_0^{\rm phys}$.
In the 3D oscillator and Coulomb benchmarks, we find $\gamma_{\rm mon}=0$ (fixed points), explaining the empirical cancellation of higher closed-period corrections; away from these integrable points, $\gamma_{\rm mon}\neq 0$ organizes genuine resurgent corrections as a trans-series controlled by Stokes automorphisms.

Now, one can reconsider the formal power series in~$Q$ with $\lambda$ as
\begin{align}
    Q_0(r) \equiv 2r\left[V(r) + \frac{\lambda^2}{2r^2} - E\right], \qquad Q_2(r) \equiv Q_2 = -\frac{1}{4} .
\end{align}
This is a renormalization improvement for $\gamma_{\rm mon}$ to vanish.
This expression gives rise to Proposition~A.1 in Ref.~\cite{Miyachi:2025ptm} if $\gamma_{\rm mon}=0$ and $\oint \magenta{dr}\, S_{\mathrm{odd}}=\hbar^{-1}\oint \magenta{dr}\, \sqrt{Q_0/r}$, such as in $3$D harmonic oscillator or Coulomb.\footnote{We would like to thank T. Miyachi and R. Namba, the authors of Ref.~\cite{Miyachi:2025ptm}, for helpful discussion. We should make it public that they and we have also derived the identical quantization condition from their open-path technique.}
In this sense of the renormalization improvement, the Langer correction is readily adopted.

\subsection{An optimized-perturbation-theory realization}
\label{sec:omp}
We now realize the above MR in an \textit{optimized/variational perturbation theory} (OPT) framework~\cite{Stevenson:1981vj,Duncan:1992ba,Buckley:1992pc,Kleinert:1995hc}, directly at the level of the exact-WKB action. \magenta{For other summation methods for perturbative quantum theory, see the review given in~Ref.~\cite{Zinn-Justin:2010zzb}.}

Let $Q(r;\lambda)$ be the radial WKB potential (with or without the Langer correction).
Let us also introduce
\begin{align}
    Q^\delta(r;\lambda) = Q^0(r;\lambda)+\delta\left[Q(r;\lambda)-Q^0(r;\lambda)\right],
    \qquad \delta\in[0,1],
\end{align}
where the seed $Q^0$ is solvable and matches the small-circle monodromy:
\begin{align}
    \text{If $r\to0$:}\quad
    Q^0(r;\lambda)\sim \frac{\lambda^2}{r^2}\qquad
    \oint_{C_0}\sqrt{Q^0} dr=2i\pi\lambda.
\end{align}
The $3$D oscillator and Coulomb are typical choices for $Q^0$.
Expand the action as a $\delta$-series at fixed parameters (omitting $dr$ \magenta{in what follows}):
\begin{align}
\oint S_{\mathrm{odd}}[Q^\delta]
=\oint \mathcal{S}^0 +\oint \delta \mathcal{S}^1+\int \delta^2 \mathcal{S}^2+\dots,
\end{align}
where
\begin{align}
    \mathcal{S}^0 = S_{\mathrm{odd}}[Q^0] = \sqrt{Q^0}, \qquad
    \mathcal{S}^1 = \frac{1}{2}\frac{Q-Q^0}{\sqrt{Q^0}},\dots\qquad
    \mathcal{S}^i = \frac{1}{i!}\frac{d^i}{d\delta^i}\sqrt{Q^\delta}, \dots .
\end{align}

Closed Voros periods are computed as (infinity–circle) minus (origin–circle), with any local divergence on $C_0$ minimally subtracted by $Z_{\rm mon}(\mu)$ so that the physical small-circle monodromy $\exp[i\pi(2l+1)]$ is maintained at every order.
Thus the centrifugal contribution of order ``$O(\hbar^2)$'' in appearance is treated as a
relevant coupling $\lambda$ (never naively expanded).

In practice, we consider a truncation order $I$, set $\delta=1$ and impose
\begin{align}
    \frac{1}{2i\pi\hbar}\oint\sum_{i=0}^{I}\mathcal{S}^i
    =n_r+\frac{1}{2} .
\end{align}
Close with the fastest apparent convergence condition, e.g.,
\begin{align}
\partial_\alpha \left[\oint S_{\mathrm{odd}}\right]_{I}=0 ,
\end{align}
on the variational couplings $\{\alpha\}$.
Solving this yields $E^{(I)}$ and the optimal $\{\alpha^{(I)}\}$.

In the case of the harmonic oscillator, we define $Q=r^2+\frac{\lambda^2}{r^2}-2E$ and
$Q^0=\alpha^2 r^2+\frac{\lambda^2}{r^2}-2E$,
\begin{equation}
\mathcal{S}^1=\frac{1}{2} \frac{(1-\alpha^2) r^2}{\sqrt{Q^0}} .
\end{equation}
Evaluating as (infinity–circle) minus (origin–circle) with MR subtraction,
\begin{align}
    \res_{r\to\infty}\mathcal{S}^0 - \res_{r=0}\mathcal{S}^0 &= \left(\frac{E}{\alpha}-\lambda\right), \\
    \res_{r\to\infty}\mathcal{S}^1 - \res_{r=0}\mathcal{S}^1 &= - \magenta{\frac{1}{2}} E\frac{1-\alpha^2}{\alpha^3} .
\end{align}
Then, the fastest apparent convergence gives
\begin{align}
    \partial_\alpha\left[\left(\frac{E}{\alpha}-\lambda\right) - \frac{1}{2}E\frac{1-\alpha^2}{\alpha^3}\right]
    = \frac{3 E (1-\alpha^2)}{2\alpha^4}= 0 ,
\end{align}
and hence $\alpha^{(1)}=1$.
Higher orders preserve $\alpha^{(I)}\to1$ and closed-period corrections cancel\magenta{\footnote{\magenta{In this case, we show that if one applies the variational method to an integrable model, the variational parameter always becomes trivial. In contrast, in Appendices we consider nonintegrable examples, where the variational parameter does not reduce to a trivial value.}}}, yielding from the leading action plus discrete phases
\begin{equation}
\frac{E}{\hbar}=2n_r+l+\frac{3}{2}.
\end{equation}
Here, note that $\oint_0^{a}=\frac{1}{2}\oint$ and $\lambda/\hbar=l+1/2$.

For the anharmonic oscillator case, the naive perturbation theory suffers from non-physical behavior of ground state energy, which should be larger than that of the harmonic oscillator, while the variational perturbation theory provides a coherent picture such that the energy is an increasing function of the corresponding deformed parameter as shown in Appendix~\ref{sec:anharmonic}.

\subsection{Some remarks}
The seed $Q^0$ (i) locks the origin monodromy via $\lambda$ and MR, and (ii) approximates the infinity behavior up to variational scales; the $\delta$-series expands only $Q-Q^0$.
The fastest appearance convergence aligns variational couplings with the true problem so that, in integrable benchmarks, closed Voros corrections beyond leading order vanish (RG fixed-point behavior).
For nonintegrable cases, the residual corrections assemble into a controlled trans-series governed by Stokes automorphisms; the MR supplies a consistent local regularization at the origin.

The open-connection formulation in $x=\ln r$ parallels boundary-to-boundary treatments of black-hole QNMs along the real axis, e.g., Andersson~\cite{Andersson:1995zk}.
Our bound-state analysis replaces the horizon ingoing condition with origin regularity, but the monodromy/connection algebra and the role of lateral Borel summation are closely
analogous.

\section{Conclusion}\label{sec:outlook}
We have provided a unified, resurgence-informed account of exact WKB quantization for radial Schr\"odinger equations with a regular singularity at the origin. The main lessons are:
\begin{enumerate}
  \item \textbf{Path choice is a gauge.} Once the nontrivial-cycle data (Voros periods, Stokes multipliers, and small-circle monodromy at $r=0$) are specified, the spectrum is independent of where one starts (turning point, punctured origin, or infinity) and of the detailed shape of the path.
  Open connection problems and closed-cycle conditions are strictly equivalent in this sense.

  \item \textbf{Origin phase vs.\ Langer bookkeeping.} The angular-momentum phase $l+\frac{1}{2}$ associated with the regular singularity can be encoded either in the momentum (Langer correction) or on the right-hand side of the quantization condition as a boundary phase; both conventions yield identical spectra when used consistently.
  It is remarkable in this sense that the $3$D harmonic oscillator is quite subtle in the exact WKB framework, while the Coulomb potential has no difficulty.
  On the other hand, at least partially, the renormalization group tells us about detailed treatments for the bookkeeping due to the relevant couplings and cutoff independence.

  \item \textbf{Two benchmark models.} For the $3$D harmonic oscillator and the $3$D Coulomb problem we carried out explicit evaluations of the closed Voros periods as (infinity-circle) minus (origin-circle) contributions, making the origin’s monodromy contribution explicit.
  Higher-order odd WKB terms integrate to zero over the relevant
  cycles, so the spectra reduce to the leading action plus discrete phases.

  \item \textbf{Boundary transfer via $r=e^x$.} The mapping $r=e^{x}$ relocates the origin's local information to the boundary $x\to-\infty$, where it becomes a convergence condition for the physical solution. This formulation makes the equivalence of open-path connection problems and closed-cycle quantization particularly transparent.

  \item \textbf{Conceptual clarification for ``physical paths''.} Our results show that the physically relevant input is the pair of boundary conditions (regular at the origin and subdominant at infinity for bound states), together with the correct Stokes data; the path itself merely implements these choices in the complex plane.
  This perspective aligns radial bound-state problems with recent open-path formulations used in QNM studies.
\end{enumerate}

The same methodology extends to multi–turning-point radial problems, to confluence limits where turning points and poles merge, and to scattering/QNM settings formulated directly in the $x=\ln r$ coordinate.
It would be interesting to quantify, beyond the integrable benchmarks studied here, when higher-order closed Voros periods cease to cancel and how the resulting corrections are organized into resurgent trans-series controlled by the underlying Stokes automorphisms.

\section*{Acknowledgements}
We are grateful to T. Miyachi and R. Namba for fruitful discussions.
We would like to thank the officers in Nishijin Plaza at Kyushu University.
This work was partially supported by Japan Society for the Promotion of Science (JSPS)
Grant-in-Aid for Scientific Research Grant Numbers
JP25K17402 (O.M.) and JP21H04975 (S.O.).
O.M.\ acknowledges the RIKEN Special Postdoctoral Researcher Program
and RIKEN FY2025 Incentive Research Projects.

\appendix

\section{Anharmonic oscillator}\label{sec:anharmonic}

The anharmonic oscillator with quartic term is given by
\begin{align}
    V(r) = \frac{1}{2} r^2 + \frac{g}{4} r^4 .
\end{align}
This potential provides the Stokes geometry as in Fig.~\ref{fig:stokes_4th-oscillator}. We find the turning points, $a_i$, as
\begin{align}
    a_1 &= \sqrt{\frac{-1+\sqrt{1+4gE}}{g}} , &
    a_2 &= i \sqrt{\frac{1+\sqrt{1+4gE}}{g}} \magenta{,} \\
    a_3 &= - \sqrt{\frac{-1+\sqrt{1+4gE}}{g}} , &
    a_4 &= - i \sqrt{\frac{1+\sqrt{1+4gE}}{g}} .
\end{align}
The dashed red curve is on another Riemann sheet through the branch cut.

\begin{figure}
\centering
\begin{tikzpicture}
    \draw[very thick] (3,0) -- (0,0) node[left] {$+$};
    \draw[very thick] (5,0) -- (8,0) node[right] {$-$};
    \draw[very thick] (5,0) -- (4,3);
    \draw[very thick] (4,3) -- (6,5) node[above] {$+$};
    \draw[very thick] (4,3) .. controls (3,3) and (2.5,3).. (1.5,5) node[above] {$-$};
    \draw[very thick] (5,0) .. controls (5.5,-3) and (6.5,-4.5).. (7,-5) node[below] {$+$};
    \draw[very thick] (3,0) .. controls (2.5,3) and (1.5,4.5).. (1,5) node[above] {$-$};
    \draw[very thick] (3,0) -- (4,-3);
    \draw[very thick] (4,-3) -- (2,-5) node[below] {$-$};
    \draw[very thick] (4,-3) .. controls (5,-3) and (5.5,-3).. (6.5,-5) node[below] {$+$};
    \draw[ultra thick,blue,snake it] (3,0) -- (4,3);
    \draw[ultra thick,blue,snake it] (5,0) -- (4,-3);
    \draw[ultra thick,red,->,rounded corners] (8,0.5) -- (5.3,0.5) -- (4,3.9) -- (2.6,0) -- (4,-3.9) -- (5.3,-0.5) -- (8,-0.5);
    \draw[ultra thick,red,dashed] (4,0) circle(0.25);
    \draw[ultra thick,red,dashed] (3.3,1.8) -- (3.8,0.2);
    \draw[ultra thick,red,dashed] (4.7,-1.8) -- (4.2,-0.2);
    \fill (5,0) circle(3pt) node [below right] {\Large $a_1$};
    \fill (4,3) circle(3pt) node [right] {\Large $a_2$};
    \fill (3,0) circle(3pt) node[below left] {\Large $a_3$};
    \fill (4,-3) circle(3pt) node [left] {\Large $a_4$};
    \draw[thick] (4,0) circle(3pt);
\end{tikzpicture}
\caption{Stokes graph for the $3$D anharmonic oscillator. $\re E>0$.}
\label{fig:stokes_4th-oscillator}
\end{figure}

The corresponding connection in the current path becomes
\begin{align}
    \begin{pmatrix}
        \varphi_l^+ \\ \varphi_l^-
    \end{pmatrix}_{\infty-i\epsilon}
    = M_{+}^\bullet \Tilde{N}_{a_4a_1} M_{+}^\bullet M_{-}^\bullet N_{a_3a_4} M_{+}^\bullet M_{-}^\bullet \Tilde{N}_{a_2a_3} M_{-}^\bullet M_{+}^\bullet N_{a_1a_2}
    \begin{pmatrix}
        \varphi_l^+ \\ \varphi_l^-
    \end{pmatrix}_{\infty+i\epsilon}
    \qquad \text{for $\epsilon>0$} .
\end{align}
Then, we have the quantization condition to vanish the coefficient of~$\varphi_l^-|_{\infty+i\epsilon}$ in $\varphi_l^+|_{\infty-i\epsilon}$
\begin{align}
    \cosh\left(\oint_{a_4}^{a_1} dr\, S_{\mathrm{odd}}\right)
    + \cos(\pi\sqrt{1+4Q_2}) = 0 .
\end{align}
Note that this form of quantization is identical to the other models which we saw above; it is reasonable that the crucial cycle is in $\re r>0$.

\begin{figure}
    \centering
    \includegraphics[width=0.7\columnwidth]{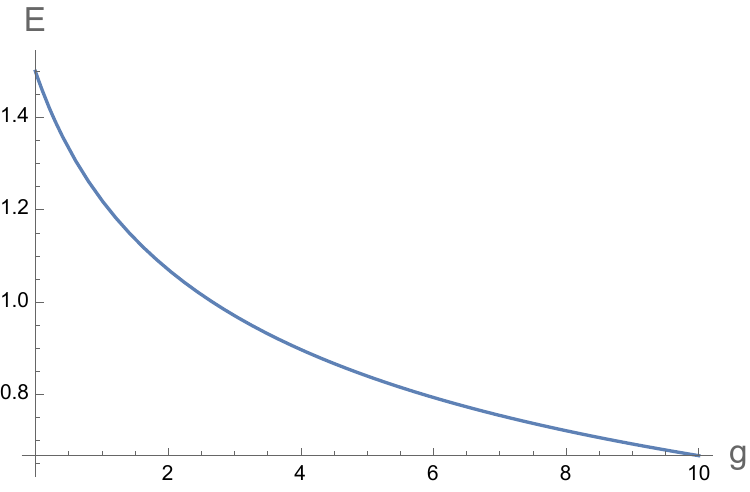}
    \caption{The ground state energy of the anharmonic oscillator as a function of~$g$. This is based on the next-to-leading estimate in the perturbation theory in terms of~$g$. The quantum numbers, $n_r$ and~$l$, are set to vanish, so $\lambda=1/2$ under the Langer correction. Typically, the energy is damping monotonically with respect to the perturbation parameter~$g$.}
    \label{fig:ho4_energy_pert}
\end{figure}

At first, the naive estimate of the next-to-leading order looks like
\begin{align}
    \oint_{a_4}^{a_1} S_{-1} dr
    &= - i \pi E - i\pi \left(l+\frac{1}{2}\right)
    -\frac{3}{16} i \pi E^2 g +
    O(g^2) .
\end{align}
Then, we have
\begin{align}
    \therefore E = \frac{4 \left(\sqrt{3 g (2n_r+l+\frac{3}{2})+4}-2\right)}{3 g} ,
\end{align}
but this energy is a monotonically decreasing function of~$g$ (see Fig.~\ref{fig:ho4_energy_pert}). For larger~$g$, the ground state energy should increase as the physical sense.

\begin{figure}[t]
    \centering
    \includegraphics[width=0.7\columnwidth]{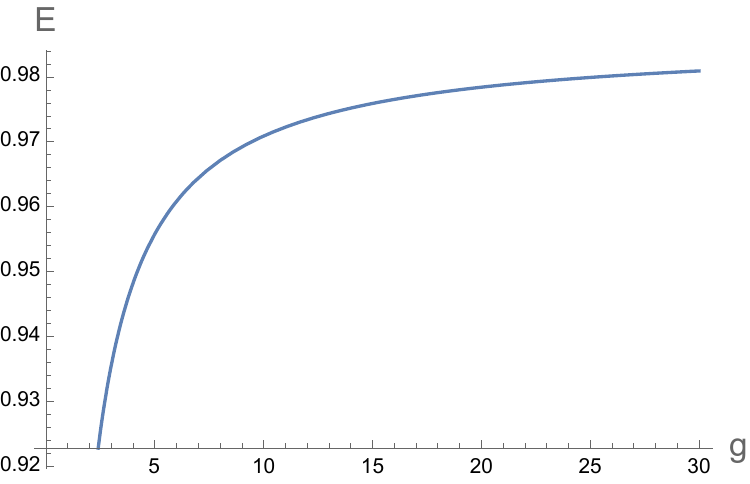}
    \caption{The ground state energy of the anharmonic oscillator as a function of~$g$. The derivation is given by the variational method with the fastest apparent condition up to subleading of~$g$, and the explicit function is given by~Eq.~\eqref{eq:ho4_energy}. The quantum numbers, $n_r$ and~$l$, are set to vanish, so $\lambda=1/2$ under the Langer correction. Typically, the energy is growing up monotonically with respect to~$g$.}
    \label{fig:ho4_energy}
\end{figure}

Now, we define $Q=r^2 + \frac{g}{2}r^4 + \frac{\lambda^2}{r^2} - 2E$ and $Q^0 = \alpha^2 r^2 + \frac{\lambda^2}{r^2} - 2 E$, and so obtain
\begin{align}
    \mathcal{S}^0 &= \sqrt{r^2 + \frac{\lambda^2}{r^2} - 2 E} , \\
    \mathcal{S}^1 &= \frac{1}{2} \frac{\frac{g}{2} r^4}{\sqrt{r^2 + \frac{\lambda^2}{r^2} - 2 E}}, \qquad \dots,
\end{align}
and
\begin{align}
    \oint \magenta{dr}\, \mathcal{S}^0 &= 2i\pi \left(\frac{E}{\alpha} - \lambda\right) ,\\
    \oint \magenta{dr}\, \mathcal{S}^1
    &= \pm i\pi \frac{\alpha^2 - E g (1 - 2 \alpha^2)}{4 g \alpha^{3}} + O(g) .
\end{align}
Note that $\mathcal{S}^1$ is expanded in terms of $g$ before integrating. The fastest apparent convergence up to $O(g)$ gives 
\begin{align}
    \partial_\alpha\left[\frac{E}{\alpha} - \lambda \pm \frac{1}{2}\frac{\alpha^2 - E g (1 - 2 \alpha^2)}{4 g \alpha^{3}}\right] = 0 ,
    \label{eq:fac_condition_anharmonic}
\end{align}
and we obtain
\begin{align}
    \alpha^{(1)} = \pm\sqrt{\frac{3gE}{1 + 10gE}}, \qquad
    \pm\sqrt{\frac{3gE}{1 - 6gE}} \magenta{,}
\end{align}
\magenta{where the first and second solutions are for $+$ and for $-$, respectively, in Eq.~\eqref{eq:fac_condition_anharmonic}.}
Hence, for instance \magenta{$\alpha^{(1)} = \sqrt{\frac{3gE}{1 + 10gE}}$}, we see
\begin{align}
    \frac{\sqrt{1 + 10gE}}{12 \sqrt{3E} g^{3/2}} + \frac{5 \sqrt{E} \sqrt{1 + 10gE}}{6 \sqrt{3g}} - \lambda &= 2n_r + 1
    \\
    \frac{(1+10 g E)^{3/2}}{\sqrt{E}} &= 12\sqrt{3}g^{3/2}(2n_r + 1 + \lambda) ,
    \label{eq:ho4_energy}
\end{align}
and depicted the ground state energy~$E$ with~$n_r=0$, $\lambda=1/2$ in Fig.~\ref{fig:ho4_energy}, which grows along~large $g$ typically.
\magenta{There are some roots of the algebraic equations with regard to~$E$. Almost those are non-physical due to, e.g., negative energy. We here illustrate a more plausible result for $g>\frac{1}{6}\sqrt{\frac{5}{2}}$.}
Of course, these are not results with such a high level of precision; even within the variational perturbation theory (with parameter $\delta$), the function of $g$ is expanded in powers of $g$, and we do not address the systematic issues of truncated fastest apparent convergence in any serious manner.

\section{Variations of Coulomb}\label{sec:cornell_yukawa}

In this Appendix, we consider the Cornell/Yukawa potential,
\begin{align}
    V_C(r) &= - \frac{\magenta{\Tilde\alpha}}{r} + \sigma r, &
    V_Y(r) &= - \frac{e^{-\mu r}}{r} ,
\end{align}
where the turning points are given by
\begin{align}
    a_C^\pm &= \frac{E}{2\sigma} \pm \sqrt{\frac{E^2}{4\sigma^2} + \frac{\magenta{\Tilde\alpha}}{\sigma}}, &
    a_Y^+ &= \frac{1}{\mu} W(-\mu/E) ,
\end{align}
respectively.
Here $W(z)$ is the Lambert W-function.
The Stokes structure of the Cornell potential is shown in Fig.~\ref{fig:stokes_cornell}. For the Yukawa potential, the Stokes structure is the same as for Coulomb potential in Fig.~\ref{fig:stokes_coulomb}.




\begin{figure}[t]
\centering
\begin{tikzpicture}
    \draw[very thick] (0,0) -- (-3,0);
    \draw[very thick] (3,0) -- (6,0) node[right] {$-$};
    \draw[very thick] (3,0) arc(40:92.5:10) node[left] {$+$};
    \draw[very thick] (3,0) arc(-40:-92.5:10) node[left] {$+$};
    \draw[very thick] (-3,0) arc(40:60:10) node[left] {$+$};
    \draw[very thick] (-3,0) arc(-40:-60:10) node[left] {$+$};
    \draw[very thick,blue,snake it] (0,0) -- (3,0);
    \fill[white] (0,0) circle(3pt);
    \draw[ultra thick] (0,0) circle(3pt) node[below] {$0$};
    \fill (3,0) circle(3pt) node[below right] {\Large $a_C^+$};
    \fill (-3,0) circle(3pt) node[below right] {\Large $a_C^-$};
    \draw[ultra thick,red,->] (6,1) -- (0,1) arc(90:270:1) -- (6,-1);
\end{tikzpicture}
\caption{Schematic Stokes graph for the Cornell potential.}
\label{fig:stokes_cornell}
\end{figure}

\begin{figure}[t]
    \centering
    \includegraphics[width=0.45\columnwidth]{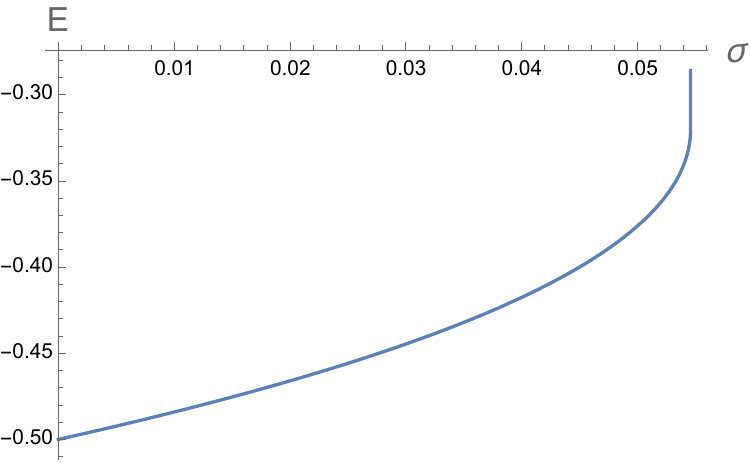}
    \hspace{1em}
    \includegraphics[width=0.45\columnwidth]{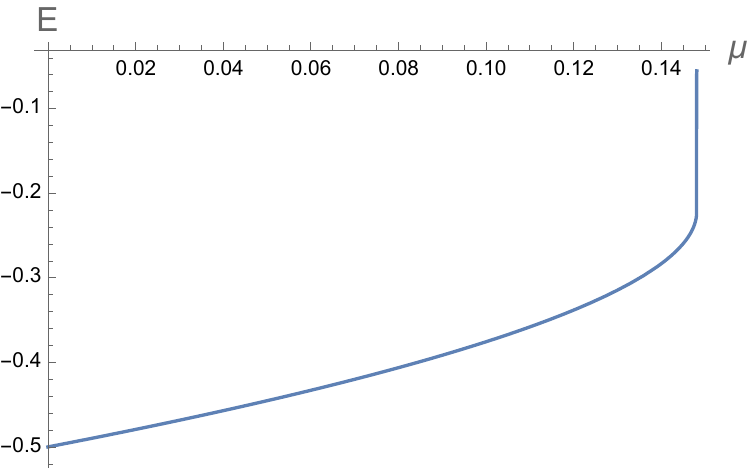}
    \caption{The ground state energy of the variations of the Coulomb potential: \textbf{(Left)} Cornell potential \magenta{with $\Tilde\alpha=1$} as a function of~$\sigma$, and \textbf{(Right)} Yukawa potential as a function of~$\mu$. The quantum numbers, $n_r$ and~$l$, are set to be zero. Typically, the energy is growing up monotonically with respect to the perturbation parameters, $\sigma$ and $\mu$.}
    \label{fig:CY_coulomb_energy}
\end{figure}

Again, we can find the same quantization condition
\begin{align}
    \cosh\left(\oint_0^{a^+} dr\, S_{\mathrm{odd}}\right)
    + \cos(\pi(2l+1)) = 0 .
\end{align}
The next-to-leading order contributes as
\begin{align}
    \frac{1}{2\pi} \oint_0^{a_C^+} dr\, S_{-1}
    &= i\frac{\magenta{\Tilde\alpha}}{\sqrt{-2E}} -\frac{3 i \magenta{\Tilde\alpha^2} \sigma }{8 \sqrt{2} (-E)^{5/2}} + O(\sigma^2), \\
    \frac{1}{2\pi} \oint_0^{a_Y^+} dr\, S_{-1}
    &= i\frac{1}{\sqrt{-2E}} -\frac{i \mu }{2 \sqrt{2} (-E)^{3/2}} + O(\mu^2) .
\end{align}
Figure~\ref{fig:CY_coulomb_energy} shows the ground state energies of the two potentials as functions of the corresponding parameters introduced to deform the Coulomb potential. The behavior in terms of $\sigma$ or $\mu$ is reasonable.

\section{Open-path $\Leftrightarrow$ Closed-cycle: a general statement}
\label{sec:open-closed}

Consider the radial Schr\"odinger equation
\begin{align}
\label{eq:radial}
    \left[- \frac{\hbar^2}{2}\frac{d^2}{dr^2} + V(r) + \frac{\hbar^2 (\ell+\tfrac12)^2}{2r^2} \right] u(r) = E\,u(r),    
\end{align}
where we have incorporated the Langer improvement $\ell(\ell+1)\to(\ell+\tfrac12)^2$.
Suppose that $V$ is meromorphic. We assume:
\begin{itemize}
\item[(A1)] $V$ extends analytically to a simply connected domain $\mathcal D\subset\mathbb C$ containing $[0,+\infty)$ and the Stokes graph relevant to the energy $E$.
\item[(A2)] $r=0$ is a \emph{regular} singular point of Eq.~\eqref{eq:radial}.
\item[(A3)] $E$ belongs to the discrete spectrum, so that a subdominant solution exists as $r\to+\infty$ in some Stokes sector.
\item[(A4)] The Borel sums of the exact-WKB objects used below exist in a fixed lateral direction $\theta$ not crossing a Stokes curve.
\end{itemize}

Let $\Gamma$ be a closed contour that goes from $+\infty$ to the vicinity of $r=0$ within a Stokes sector, encircles $r=0$ once counterclockwise on the same sheet, and returns to $+\infty$.

\begin{prop*}[\textbf{Open-path $\Leftrightarrow$ Closed-cycle under A1--A4}]
For a given $E$, the following are equivalent:
\begin{description}
\item[(i)] \textbf{Open-path quantization:} There exists a nontrivial solution of Eq.~\eqref{eq:radial} which is subdominant as $r\to+\infty$ and \emph{regular} at $r=0$ (i.e. no logarithmic growth in the Frobenius expansion).
\item[(ii)] \textbf{Closed-cycle exact-WKB quantization:} The monodromy gathered along $\Gamma$ has a zero point, i.e.
\begin{align}
    \label{eq:closed-quantization}
    \cosh\left(\oint_\Gamma dr\, S_{\mathrm{odd}}\right)
    + \cos(\pi\sqrt{1+4Q_2}) = 0 .
\end{align}
In other words, the monodromy phase in terms of~$\gamma_{\mathrm{mon}}$ has a unit with the ordered product of Stokes multipliers encountered along $\Gamma$.
\end{description}
\end{prop*}

\paragraph{Sketch of proof}
Fix a decaying solution $\psi_\infty$ at $+\infty$ (lateral Borel sum at angle $\theta$), and a local Frobenius basis $(\psi^{\mathrm{reg}}_0,\psi^{\mathrm{sing}}_0)$ at $r=0$, where $\psi^{\mathrm{reg}}_0$ and $\psi^{\mathrm{sing}}_0$ are the regular and singular solutions at the origin, respectively. Transport $\psi_\infty$ to the origin across Stokes curves, multiplying by the corresponding monodromy matrices (AKT connection). One obtains
\begin{align}
    \psi_\infty = C_{\mathrm{reg}}(E,\hbar)\,\psi^{\mathrm{reg}}_0 + C_{\mathrm{sing}}(E,\hbar)\,\psi^{\mathrm{sing}}_0,
\end{align}
with connection coefficients expressed through Voros and Stokes multipliers. Condition (i) in Proposition holds iff $C_{\mathrm{sing}}(E,\hbar)=0$. Now close the path by a small counterclockwise loop around $r=0$ and return to $+\infty$; the net transport along this closed contour amounts to multiplication by Stokes multipliers and~$e^{2\pi i\gamma_{\mathrm{mon}}}$ in a suitable basis. Triangularity of the local monodromy at a regular singularity and Wronskian conservation (i.e., the condition so that the coefficient of dominant solution on the right hand side should vanish as usual) imply that $C_{\mathrm{sing}}=0$ iff the closed monodromy has a zero point (monodromy unit), which is precisely Eq.~\eqref{eq:closed-quantization}. Independence of Eq.~\eqref{eq:closed-quantization} from local normalizations follows from the standard arbitrariness of choosing Stokes data.
\hfill$\square$

For the 3D harmonic oscillator and for the Coulomb problem, one can choose $\Gamma$ such that $\gamma_{\mathrm{mon}}=0$; then the second term on the left-hand side in Eq.~\eqref{eq:closed-quantization} is moved to the right-hand side (with the Langer improvement).

The statement presumes regularity at the origin and excludes irregular singular points and energies for which Stokes curve accumulate around $r=0$.
If $V$ has additional singularities or branch points in $\mathcal D$, the contour $\Gamma$ must be chosen to avoid them (see Appendix~\ref{sec:anharmonic}); otherwise extra factors contribute.

The Bohr--Sommerfeld quantization is now given in the form as
\begin{equation}
\label{eq:QC-practical}
\frac{1}{2\pi\hbar}\oint_\Gamma p_{\mathrm{eff}}(r;E)\,dr
\;+\; \Phi_\Gamma(E,\hbar) \;+\; \gamma_{\mathrm{mon}}(E,\hbar)\;\in\; \mathbb Z,
\end{equation}
where
$p_{\mathrm{eff}}(r;E):=\sqrt{2m\,(E-V_{\mathrm{eff}}(r))}$ with $V_{\mathrm{eff}}(r):=V(r)+\hbar^2(\ell+\tfrac12)^2/(2r^2)$, and $\Phi_\Gamma$ collects all higher-order WKB/Voros corrections and Stokes contributions along $\Gamma$.

\bibliographystyle{utphys}
\bibliography{ref,ref_ewkb,ref_res}
\end{document}